\documentclass[10pt]{article}

\usepackage{amssymb,amsfonts,amsmath}
\usepackage{enumerate,float}
\usepackage{textcomp}
\usepackage{color}
\usepackage{tcolorbox,pict2e}
\usepackage{authblk}
\usepackage{tikz}\usetikzlibrary{tikzmark}\usetikzlibrary{calc}
\usetikzlibrary{arrows,snakes,backgrounds}
\usetikzlibrary{calc, knots, hobby, arrows.meta}
\pgfdeclarelayer{background layer}
\pgfdeclarelayer{foreground layer}
\pgfsetlayers{background layer,main,foreground layer}
\tikzstyle{arrow} = [thick,-{Stealth[length=5mm]}]

\usepackage[natbib=true, backend = bibtex, style=numeric-comp, sorting=none]{biblatex}
\def\be{\begin{eqnarray}}
\def\ee{\end{eqnarray}}

\def\Tr{{\rm Tr}\,}

\definecolor{red}{rgb}{1,0,0}
\definecolor{orange}{rgb}{1,0.5,0}
\definecolor{violet}{rgb}{0.7,0,1}

\hoffset= - 1.0in         

\begin{document}

\title{\bf Can Yang-Baxter imply Lie algebra?}

\author[1,2]{{\bf D. Khudoteplov}\thanks{{khudoteplov.dn@phystech.edu}}}
\author[2,3]{{\bf A. Morozov}\thanks{{morozov@itep.ru}}}
\author[1,2,3]{{\bf A. Sleptsov}\thanks{{sleptsov@itep.ru}}}

\vspace{4.5cm}

\affil[1]{Moscow Institute of Physics and Technology, 141700, Dolgoprudny, Russia}
\affil[2]{Institute for Information Transmission Problems, 127051, Moscow, Russia}
\affil[3]{NRC "Kurchatov Institute", 123182, Moscow, Russia\footnote{former Institute for Theoretical and Experimental Physics, 117218, Moscow, Russia}}
\renewcommand\Affilfont{\itshape\small}

\date{}

\maketitle

\vspace{-6.8cm}

\begin{center}
	\hfill ITEP/TH-09/25\\
	\hfill IITP/TH-07/25\\
	\hfill MIPT/TH-07/25
\end{center}

\vspace{4.8cm}

\centerline{ABSTRACT}

\bigskip

{\footnotesize
Quantum knot invariants (like colored HOMFLY-PT or Kauffman polynomials) are a distinguished class of non-perturbative topological invariants. 
Any known way to construct them (via Chern-Simons theory or quantum R-matrix) starts with a finite simple Lie algebra. Another set of knot invariants -- of finite type -- is related to quantum invariants via a perturbative expansion. However can all finite type invariants be obtained in this way? Investigating this problem, P. Vogel discovered a way to polynomially parameterize the expansion coefficients with three parameters so that, at different specific values, this reproduces the answers for all simple Lie (super)algebras.
Then it is easy to construct a polynomial $P_{alg}$ that vanishes for all simple Lie algebras, and the corresponding Vassiliev invariant would thus be absent from the perturbative expansion.

We review these Vogel claims pointing out at least two interesting implications of his construction. First, we discuss whether infinite-dimensional Lie algebras might enlarge Chern-Simons theory. Second,  Vogel's construction implies an alternative axiomatization of simple Lie algebras -- when we start from knot invariants and arrive at Lie algebras and their classification, which is opposite to conventional logic that we mentioned at the beginning.

}

\bigskip

\section{Introduction}

One of the tasks of modern theory is to characterize (if not calculate)
the gauge-invariant observables in Yang-Mills theory, which are basically the averages
of Wilson lines.
So far the best achievement is identification of these quantities in the simplest
3d Chern-Simons theory with knot polynomials.
Since the theory is topological, Wilson averages depend only on the linking of contours
and can be analyzed by methods, alternative to conventional quantum field theory (QFT) --
what allows to drastically enrich QFT methods and obtain exact non-perturbative answers. The basic ingredient in this approach is the Yang-Baxter equation, which is one of the Reidemeister equivalences -- and in Yang-Mills QFT it appears through the quantization of the underlying Lie algebra and associated universal $R$-matrices. This is known as the Reshetikhin-Turaev formalism. 

Within this approach it is not at all obvious that all topological invariants are
described by Wilson averages, i.e. by knot polynomials. Clearly this is extremely important for QFT -- to know if there can be (non-perturbative) observables, different from Wilson averages. Among many attempts to resolve this puzzle, distinguished is the approach, suggested by P.Vogel \cite{vogel2011algebraic}, because it sounds conceptually very simple. Namely, he proposed to construct a kernel of a mapping (called a weight system) from the universal Vassiliev invariant (called the Kontsevich integral \cite{kontsevich1993vassiliev}) to the knot polynomials using a universal description of the adjoint representation of the Lie algebras. The claim is that a representation theory of Lie algebras allows a new kind of universal description, where all Dynkin diagrams associate with concrete values of three parameters ($\alpha, \beta, \gamma$) and many quantities in Lie algebra, from dimensions to knot polynomials \cite{vogel1999universal, mkrtchyan2012casimir, Mkrtchyan2012qdim, mironov2016universal, mironov2016racah, isaev2023cube}, are expressed analytically through these parameters.

More specifically, it is known \cite{birman1993knot, bar1995vassiliev} that any coefficient of the perturbative expansion of the Wilson average in the Chern-Simons theory for any knot $\mathcal{K}$, a finite gauge algebra $\mathfrak{g}$ and its finite-dimensional representation $R$ is a some Vassiliev invariant (i.e. invariant of the finite type) \cite{vassiliev1990cohomology}. It gives a perturbative interpretation of the Vassiliev theory within the Chern-Simons framework. For this reason, it has long been considered that the study of non-perturbative invariants (knot polynomials) is a priority, since they contain more information (or at least no less) than perturbative ones.

However, the work of P. Vogel cast doubt on this point of view. He proved that there exist such Vassiliev invariants that cannot be obtained as coefficients of a perturbative expansion. Moreover, he presented one such invariant as a dual element to explicitly constructed closed Jacobi diagram. This allows us to take a new look at the old discussion about the existence of other non-perturbative observables besides Wilson loops.

\medskip

The possibility of a universal description of the representation theory of Lie algebras (even of the adjoint sector) deserves special attention. Vogel himself had an idea to construct a universal Lie algebra \cite{vogel1999universal} which has not yet been implemented due to zero divisors. He introduced a peculiar $\Lambda$-algebra of (equivalence classes of) the 3-valent diagrams to describe Reidemeister-invariant quantities, and found a map from the polynomials of $\alpha$, $\beta$, $\gamma$ into it. This map has a kernel, which contains 
\be
\label{pclassic}
P_{alg} = P_{sl_2} P_{sl}P_{so}P_{exc} P_{D_{2,1}},
\ee
multiples of $ P_{alg} $ and potentially some other undiscovered polynomials.

Note that $ P_{alg} $ vanishes for all the simple Lie algebras and superalgebras, because each factor individually vanishes for the corresponding algebra or series of algebras. In other words, there is no polynomial that is not associated with a Lie algebra, which has a counterpart in the space of invariant 3-valent diagrams. We describe this statement and discuss its implications.

Honestly speaking, this is not an exhausting argument, there are at least
three loopholes:
\begin{itemize}

\item Vogel takes into account that between Vassiliev invariants there can only be polynomial (algebraic) relations. However, nothing forbids the existence of more complicated (transcendental) relations.

\item Instead of a finite gauge Lie algebra, one can consider an infinite-dimensional one. Such Chern-Simons theories have not been formulated to date, since a correct definition of a trace and, in general, a contraction of indices between structure constants is required.

\item Also one can consider infinite-dimensional representations of classical algebras. Such representations may lead to divergent contributions, which would require additional specifications or regularizations, which could potentially help.

\end{itemize}

Still the story deserves to be widely known to physical audience and further studied, understood and interpreted.

\bigskip

Our paper is organized as follows. In Section \ref{sec:2} we briefly review the Chern-Simons theory and the Kontsevich integral. Then we formulate the problem of whether all Vassiliev invariants are contained in the Chern-Simons invariants. Next we briefly review Vogel's approach via $\Lambda$-algebra, which allows one to solve this problem. In Section \ref{sec:3} we discuss three implications of Vogel's result. First, we discuss whether it is possible to distinguish all knots using Chern-Simons theory with a finite gauge Lie algebra. Second, we discuss the infinite-dimensional case. Third, we note that if we invert Vogel's logic, we can obtain an alternative axiomatization and classification of Lie algebras. Namely, starting from knot theory, we can naturally arrive at the notion of a Lie algebra as the dual space of a Vogel algebra, and the zero divisor in this algebra gives the Cartan-Dynkin classification. In summary (Section \ref{sec:4}) we collect all interesting results and open problems, which we discuss in this paper.

\section{On the kernel of Lie algebra weight system}
\label{sec:2}
\subsection{Chern-Simons theory and quantum knot invariants}

The 3-dimensional Chern-Simons action with gauge group $G$ for vector field $A_{\mu} = A_{\mu}^a(x)\,T^a$ ($T^a$ are generators of the corresponding Lie algebra $\mathfrak{g}$) is
\be
S(A) = \dfrac{\kappa}{4\pi}\int_{S^3}d^3x \,\epsilon^{\mu\nu\rho}\,\Tr\left(A_{\mu}\partial_{\nu}A_{\rho} + \frac{2}{3}A_{\mu}A_{\nu}A_{\rho}\right).
\ee

Gauge invariant functions of $A_{\mu}$ are Wilson loops
\be\label{wilson}
\langle W_R(\mathcal{K}) \rangle = \dfrac{1}{Z} \int \mathcal{D}A \ {\rm e}^{iS(A)} \ \Tr_R\left( {\rm Pexp} \ i\oint_CA_{\mu}dx^{\mu} \right),
\ee
that determine topological invariants of contour $C$. Partition function $Z$ is defined as usual $Z = \int \mathcal{D}A \ {\rm e}^{iS(A)}$. These functions \eqref{wilson} are known as quantum knots invariants or Witten-Reshetikhin-Turaev invariants \cite{Witten1988hf, Reshetikhin:1990pr, reshetikhin1991invariants}. They equal to invariants calculated with the help of quantum R-matrices in the corresponding representation of quantum algebra $U_q(\mathfrak{g})$. 

If we apply the standard perturbation technique to Wilson loops
\begin{enumerate}
	\item Wick's rotation $A_{z}^{a}=A_{1}^{a}-i A_{2}^{a}, \  A_{\bar{z}}^{a}=A_{1}^{a}+i A_{2}^{a}$;
	\item gauge fixing $A_{\bar{z}}=0$;
	\item quadratic action $\left.S(A)\right|_{A_{\bar{z}}=0}=i\,\displaystyle\int dt d\bar{z} dz\,\epsilon^{m n} A_{m}^{a} \partial_{\bar{z}} A_{n}^{a}$;
	\item pair correlator
	$
	\label{lcp}
	\langle\, A_{m}^{a}(t_{1},z_{1},\bar{z}_{1})\, A_{n}^{b}(t_{2},z_{2},\bar{z}_{2}) \,\rangle = \epsilon_{m n} \,\delta^{a b}\, \dfrac{\hbar}{2\pi i}\dfrac{\delta(t_{1}-t_{2})}{z_{1}-z_{2}}$ with $\hbar=\frac{2\pi i}{\kappa + N}$;
	\item Wick theorem,
\end{enumerate}
then we get the following answer
{\small
	\be
	\label{wilsonhol}
	\langle\, W^{\mathfrak{g}}_R(\mathcal{K})\,\rangle=\sum\limits_{n=0}^{\infty} \,\dfrac{\hbar^n}{(2\pi i)^n}\,\displaystyle\int\limits_{o(z_{1})<o(z_{2})<...<o(z_{n})}\,\sum\limits_{p\,\in P_{2n}}\,(-1)^{p_\downarrow}\,
	\bigwedge\limits_{k=1}^{n}\, \dfrac{dz_{i_k}-dz_{j_k} }{z_{i_k}-z_{j_k}}\cdot {\Tr}_{R}\Big( T^{a_{\sigma_{p}(1)}} T^{a_{\sigma_{p}(2)}}...T^{a_{\sigma_{p}(2n)}} \Big),
	\ee
}
where the integration is over the contour $\mathcal{K}$ with the given orientation $o$; the sum runs over the set of all pairing $P_{2n}$ of $2n$ numbers, an element of this set has the form $p=((i_1,j_1)...(i_n,j_n))$ where $i_k < j_k$ and the numbers $i_k,j_k$ are all different numbers from the set $\{1, 2, . . . , 2n\}$. A function $\sigma_p$ returns a minimum number from the pair $(i_k , j_k)$. A symbol $p_\downarrow$ denotes the number of down-oriented segments between critical points on the knot $\mathcal{K}$ entering the integral. More details one can find in \cite{labastida1998kontsevich, sleptsov2014hidden}.

From this expansion we see that the information about the knot and the gauge
group $\langle\, W_R(\mathcal{K})\,\rangle$ contributes separately. The information about the embedding of knot into $S^3$ is encoded in the integrals of the form:
$$
v_{n,m}^{\mathcal{K}} \sim \int\limits_{o(z_{1})<...<o(z_{n})}\,\sum\limits_{p\,\in P_{2n}}\,(-1)^{p_\downarrow}\,
	\bigwedge\limits_{k=1}^{n}\, \dfrac{dz_{i_k}-dz_{j_k} }{z_{i_k}-z_{j_k}}
$$
and the information about the gauge group and representation enter the answer as the group factors:
$$
G_{n,m}^{\mathfrak{g},R} \sim {\Tr}_{R}\Big( T^{a_{\sigma_{p}(1)}} T^{a_{\sigma_{p}(2)}}...T^{a_{\sigma_{p}(2n)}} \Big).
$$

Thus we can rewrite the perturbative series \eqref{wilsonhol} for the Wilson loop in a more schematic way
\begin{equation}
\label{wilsonexpansion}
\langle\, W^{\mathfrak{g}}_R(\mathcal{K})\,\rangle = \sum_{n=0}^{\infty} \hbar^n \sum_{m=1}^{\mathcal{N}_n} G_{n,m}^{\mathfrak{g},R} \cdot v_{n,m}^{\mathcal{K}},
\end{equation}
where $\mathcal{N}_n$ is the number of independent group elements of degree $n$.

Coefficients $v_{n,m}^{\mathcal{K}}$ are Vassiliev invariants of order $n$. Detailed description of Vassiliev invariants and the corresponding algebra can be found in textbook \cite{chmutov2012introduction}. Also, the necessary information in a much more concise form is contained in Section 2 in paper \cite{khudoteplov2024construction}.

Reshetikhin-Turaev formalism is related to the $A_0=0$ gauge \cite{guadagnini1990braids,guadagnini1990chern,morozov2010chern}, but the gauge used by Kontsevich $A_{\bar z}=0$, although less restrictive (leaving holomorphic integrals instead of individual points), turns out to be better justified. The \textit{perturbative} Vassiliev invariants obtained from it may turn out to be richer than the \textit{non-perturbative} colored HOMFLY-PT polynomials etc\footnote{"etc" refers to the analogues of colored HOMFLY-PT for other algebras, say, to the Kauffman polynomials for the $so$ and $sp$ series, and their still under-investigated analogues for exceptional groups.}, but this is still unclear. Literally, the perturbative HOMFLY-PT etc decompositions reproduce only \textbf{some} combinations of Vassiliev invariants, and whether it is possible or not to reproduce all of them separately is unknown.	


\subsection{Chern-Simons invariants and the Kontsevich integral}
\label{CSKI}
In 1993 it was proved that any coefficient of a quantum knot invariant is a Vassiliev invariant \cite{birman1993knot, bar1995vassiliev}. It is natural to ask whether the converse is true. This question was formulated by D.Bar-Natan in his 1995 paper \cite{bar1995vassiliev}: \emph{Are all Vassiliev invariants contained in quantum knot invariants considered  for all (semi)simple Lie (super)algebras $\mathfrak{g}$}? P.Vogel proved the existence of the kernel of any Lie algebra weight systems what imply the negative answer to the Bar-Natan's question \cite{vogel2011algebraic}. To discuss Vogel's answer to this question in details, it is necessary to reformulate it in more appropriate terms. To do this, one should think about the Kontsevich integral as a generating series for all Vassiliev invariants (i.e. the universal Vassiliev invariant).

A knot invariant $v$ is called a Vassiliev invariant of order no more than $n$ if $v$ vanishes at singular knots with $\geq n+1$ double points. The continuation of knot invariants to the set of singular knots is determined by \textit{Vassiliev skein relation}:
\begin{figure}[h]
	\centering
	\begin{tikzpicture}[scale=0.4]
		\coordinate (a) at (-7, 0);
		\coordinate (b) at (-0.5, 0);
		\coordinate (c) at (6, 0);
		\begin{pgfonlayer}{background layer}
			\draw [-{Stealth[length=3mm]}, ultra thick] ($(c)-(1, 1)$) -- ($(c)+(1, 1)$);
			\draw [-{Stealth[length=3mm]}, ultra thick] ($(b)-(-1, 1)$) -- ($(b)+(-1, 1)$);
			\draw [-{Stealth[length=3mm]}, ultra thick] ($(a)-(1, 1)$) -- ($(a)+(1, 1)$);
			\draw [-{Stealth[length=3mm]}, ultra thick] ($(a)-(-1, 1)$) -- ($(a)+(-1, 1)$);
		\end{pgfonlayer}
		\begin{pgfonlayer}{main}
			\draw [white, line width=10pt] ($(c)+(1, -1)$) -- ($(c)+(-1, 1)$);
			\draw [-{Stealth[length=3mm]}, ultra thick] ($(c)+(1, -1)$) -- ($(c)+(-1, 1)$);
			
			\draw [white, line width=10pt] ($(b)+(-1, -1)$) -- ($(b)+(1, 1)$);
			\draw [-{Stealth[length=3mm]}, ultra thick] ($(b)+(-1, -1)$) -- ($(b)+(1, 1)$);
			\draw [fill] (a) circle(0.15);	
		\end{pgfonlayer}
		\begin{pgfonlayer}{foreground layer}
			\draw [dashed] (a) circle (1.4142);
			\draw [dashed] (b) circle (1.4142);
			\draw [dashed] (c) circle (1.4142);
			\node [scale =1.6] at ($(a)+(-2.4, 0)$) { \textit{v} } ;
			\node [scale =2.5] at ($(a)+(-1.7, 0)$) { ( } ;
			\node [scale =2.5] at ($(a)+(1.7, 0)$) { ) } ;
			\node [scale =1.6] at ($(b)+(-2.4, 0)$) { \textit{v} } ;
			\node [scale =2.5] at ($(b)+(-1.7, 0)$) { ( } ;
			\node [scale =2.14] at ($(b)+(1.7, 0)$) { ) } ;
			\node [scale =1.6] at ($(c)+(-2.4, 0)$) { \textit{v} } ;
			\node [scale =2.5] at ($(c)+(-1.7, 0)$) { ( } ;
			\node [scale =2.5] at ($(c)+(1.7, 0)$) { ) } ;
			\node [scale=1.8] at ($0.6*(b) + 0.4*(c)+(0.2, 0)$) {$-$};
			\node [scale=1.8] at ($0.6*(a) + 0.4*(b)+(0.2, -0.05)$) {$=$};
		\end{pgfonlayer}
	\end{tikzpicture}
	\vspace{-3mm}
	\label{fig:Vass-skein}
\end{figure}

For our aims the rigorous definition of these invariants is not necessary. It is important that the Vassiliev invariants satisfy certain relations called $1T$ and $4T$:
\begin{figure}[H]
	\vspace{-2mm}
	\centering\leavevmode
	\includegraphics[width=9.5cm]{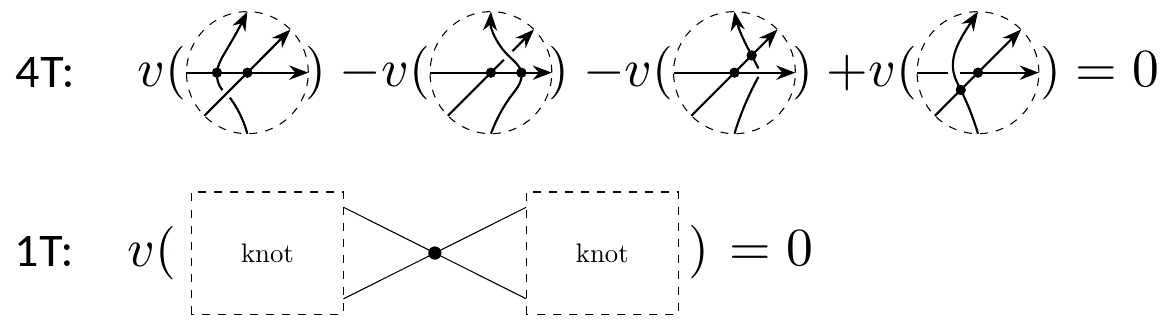}
	\vspace{-3mm}
\end{figure}
Also it is important that the Vassiliev invariants of order $n$ form an algebra $\mathcal{V}_n$ with respect to following multiplication. The product of two Vassiliev invariants of orders $\leq n$ and $\leq m$ is a Vassiliev invariants of order $\leq n+m:$ \ $v_n\cdot v_m = v_{n+m}$. It provides polynomial relations on them.

Invariants of finite type have a combinatorial description in terms of chord diagrams. A chord diagram of order $n$ is an oriented circle with $n$ pairs of distinct points. The set of chord diagrams of order $n$ is denoted $\mathbf{A}_n$. Algebra of chord diagrams is a vector space $\mathcal{A}_n = \large\textbf{A}_n/\langle\text{4T, 1T} \rangle$  modulo the four-term and the one-term relations, corresponding to the relations on the Vassiliev invariants,	with a well-defined multiplication:
\begin{figure}[H]
	\centering
\begin{tikzpicture}[scale=0.5]
\put(-70,-3){$4T =$}
\coordinate (a) at (-2, 0);
\coordinate (b) at (1, 0);
\coordinate (c) at (4, 0);
\coordinate (d) at (7, 0);
\draw[dashed] (a) circle (1);
\draw[thick] ($(a)+(240:1)$) arc(240:300:1);
\draw[thick] ($(a)+(0:1)$) arc(0:60:1);
\draw[thick] ($(a)+(140:1)$) arc(140:180:1);
\draw ($(a)+(255:1)$)--($(a)+(160:1)$);
\draw ($(a)+(280:1)$)--($(a)+(20:1)$);

\draw[dashed] (b) circle (1);
\draw[thick] ($(b)+(240:1)$) arc(240:300:1);
\draw[thick] ($(b)+(0:1)$) arc(0:60:1);
\draw[thick] ($(b)+(140:1)$) arc(140:180:1);
\draw ($(b)+(285:1)$)--($(b)+(160:1)$);
\draw ($(b)+(255:1)$)--($(b)+(20:1)$);

\draw[dashed] (c) circle (1);
\draw[thick] ($(c)+(240:1)$) arc(240:300:1);
\draw[thick] ($(c)+(0:1)$) arc(0:60:1);
\draw[thick] ($(c)+(140:1)$) arc(140:180:1);
\draw ($(c)+(45:1)$)--($(c)+(160:1)$);
\draw ($(c)+(270:1)$)--($(c)+(20:1)$);

\draw[dashed] (d) circle (1);
\draw[thick] ($(d)+(240:1)$) arc(240:300:1);
\draw[thick] ($(d)+(0:1)$) arc(0:60:1);
\draw[thick] ($(d)+(140:1)$) arc(140:180:1);
\draw ($(d)+(20:1)$)--($(d)+(160:1)$);
\draw ($(d)+(270:1)$)--($(d)+(45:1)$);

\node at ($0.5*(a)+0.5*(b)$) {$-$};
\node at ($0.5*(c)+0.5*(b)$) {$-$};
\node at ($0.5*(c)+0.5*(d)$) {$+$};
\end{tikzpicture} \qquad \qquad
\begin{tikzpicture}[scale=0.5]
\put(47,10){$,$}
\put(57,10){$1T =$}
\draw[dashed] (7, 1) circle (1);
\draw[thick] (7, 1)+(160:1) arc(160:200:1);
\draw[thick] (7, 1)+(340:1) arc(340:380:1);
\draw (6,1)--(8,1);
\draw[dotted] ($(7, 1)+(160:1)$) --($(7, 1)+(20:1)$);
\draw[dotted] ($(7, 1)+(200:1)$) --($(7, 1)+(340:1)$);
\end{tikzpicture}
\end{figure}
\hspace{-5mm}
\begin{equation*}
\begin{picture}(850,5)(-380,-35)
\put(-280,-23){multiplication:}
\put(-178,-20){\circle{32}}
\put(-194,-20){\line(1,0){32}}
\put(-178,-36){\line(0,1){32}}
\put(-158,-22){$\cdot$}
\put(-143,-34){\line(0,1){28}}
\put(-127,-34){\line(0,1){28}}
\put(-135,-36){\line(0,1){32}}
\put(-151,-20){\line(1,0){32}}
\put(-135,-20){\circle{32}}
\put(-108,-22){$=$}
	
\put(-90,-10){\line(1,-3){9}}
\put(-90,-30){\line(1,3){9}}
	
\put(-74,-2){\line(1,-2){14}}
\put(-65,-5){\line(-1,-3){11}}
\put(-59,-12){\line(-1,-3){8}}
\put(-62,-8){\line(-1,-3){10}}
\put(-75,-20){\circle{36}}
\end{picture}
\label{anmult}
\end{equation*}
This algebra is isomorphic to the algebra $\mathcal{A}_n$ of chord diagrams with $n$ chords (see details in \cite{chmutov2012introduction, khudoteplov2024construction}). More precisely, let $\mathcal{W}_n = \mathcal{A}_n^{*} = \text{Hom}~(\mathcal{A}_n, \mathbb{R})$ be a space of linear functions on $\mathcal{A}_n$. Then $\mathcal{W} = \bigoplus_{n=0}^{\infty} \mathcal{W}_n \cong \bigoplus_{n=0}^{\infty} \mathcal{V}_n/\mathcal{V}_{n-1} = \mathcal{V}$. We choose a basis $D_{n,m}$ in $\mathcal{A}_n$ and its dual basis $v_{n,m}$ and consider the following generating series:
\begin{equation}
I(\mathcal{K}) = \sum_{n=0}^{\infty} \hbar^n \sum_{m=1}^{\dim {\mathcal{V}}_n} D_{n,m} \cdot v_{n,m}^{\mathcal{K}}.
\end{equation}
This is the celebrated Kontsevich integral.

What is the connection with the invariants from the Chern-Simons theory? Choose a basis $T^1,.., T^r$ of $\mathfrak{g}$ and let $T_1,.., T_r$ be the dual basis with respect to the non-degenerate bilinear form. Then we consider irreducible finite-dimensional representation $R$ of universal enveloping algebra $U(\mathfrak{g})$ and define the function $\varphi^{\mathfrak{g}}_{R}$ on chord diagrams, which satisfies $4T$ relation:
\be
\label{weightsys}
\varphi^{\mathfrak{g}}_{R}: \mathcal{A}_n \rightarrow ZU(\mathfrak{g}) \rightarrow \mathbb{C}, \quad D \mapsto \Tr_R \left( T^aT^b... \right)
\ee
\begin{picture}(850,30)(-141,-33)
\put(-35,-22){$D_{2,1}=$}
\put(0.5,-12){$*$}
\put(32,-22){$ \ , \quad \varphi_{R}^{\mathfrak{g}}(D_{2,1})=\displaystyle \Tr_R\left( \sum_{a,b,c=1}T^a\, T^b\, T^c\, T_{b}\, T_{a}\, T_{c} \right)$}
\put(7,-34){\line(0,1){28}}
\put(23,-34){\line(0,1){28}}
\put(-1,-20){\line(1,0){32}}
\put(15,-20){\circle{32}}
\put(7,-6.5){\circle*{2}}
\put(23,-6.5){\circle*{2}}
\put(31,-20){\circle*{2}}
\put(1,-4){\mbox{\fontsize{7.5}{7.5} $a$}}
\put(28,-18){\mbox{\fontsize{7.5}{7.5} $c$}}
\put(18,-4){\mbox{\fontsize{7.5}{7.5} $b$}}
\end{picture}

\

The definition of function $\varphi^{\mathfrak{g}}_{R}$ should be clear from the above example for chord diagram $D_{2,1}$. Applying this function to the Kontsevich integral we obtain the Wilson loop expectation value for the Chern-Simons, what can be derived in a (anti)holomorphic gauge as was discussed in the previous section:

\vspace{-5mm}	
\be
\varphi^{\mathfrak{g}}_{R} \Big( I(\mathcal{K}) \Big) = \langle\, W^{\mathfrak{g}}_R(\mathcal{K})\,\rangle.
\ee

Thus, we can reformulate Bar-Natan's question in terms of function $\varphi$: For which algebras $\mathfrak{g}$ and their representations $R$ the kernel of $\varphi$ is empty
\be
\text{Ker} \left(\, \varphi^{\mathfrak{g}}_{R} \, \right) = \, \varnothing \, ?
\ee

It is easy to see that for any representations of algebra $sl_2$ this kernel is nontrivial. Indeed, by definition \eqref{weightsys} $\varphi^{\mathfrak{g}}_{R}$ is an eigenvalue of the $n$th Casimir operator corresponding to $\varphi^{\mathfrak{g}}\left(D_{n,m}\right)$, while in $sl_2$ case independent Casimir operators are the second one $C_2$ and it powers $C_2^2, C_2^3, \ldots$ Therefore, on the 4th order for any representation of $sl_2$ there are only two independent group factors $C_2(R)$ and $C_2^2(R)$, while independent diagrams are three $D_{2,1}^2$, $D_{4,2}$ and $D_{4,3}$. Thus, $\dim\Big( \text{Ker} \big(\, \varphi^{sl_2}_{R} \,\big) \Big) = 1$. 

Moreover, $\text{Ker} \left(\, \varphi^{\mathfrak{g}}_{R} \, \right) \neq \, \varnothing$ for any particular simple algebra $\mathfrak{g}$. Therefore, the correct question is the following: Do there exist such finite (semi)simple Lie (super)algebras $\mathfrak{g}_1$, $\mathfrak{g}_2$, ..., $\mathfrak{g}_k$ such that 
\be
\label{question}
\text{Ker} \left(\, \varphi^{\mathfrak{g}_1}_{R_1} \, \right) \cap \text{Ker} \left(\, \varphi^{\mathfrak{g}_2}_{R_2} \, \right) \cap \ldots \cap  \text{Ker} \left(\, \varphi^{\mathfrak{g}_k}_{R_k} \, \right) = \, \varnothing \, ?
\ee

P.Vogel answered to this question and this answer is negative.

\subsection{Constructing the kernel}
\subsubsection{Algebra of closed Jacobi diagrams}
\label{Jacobialg}
Now let us review Vogel's approach to construct a particular closed Jacobi diagram, which is not determined by any finite-dimensional Lie weight system. The algebra of chord diagrams $\hat{\mathcal{A}}_n$ is isomorphic to the algebra of closed Jacobi diagrams $\hat{\mathcal{C}}_n$, which is the vector space of closed diagrams with only trivalent $2n$ vertices and distinguished cycle modulo STU and 1-term relations:

\begin{figure}[H]
	\centering
	\begin{tikzpicture}[scale=0.6]
	\draw[ultra thick]  (-2, 0) circle (1);
	\draw (-2, 0) -- ($(-2, 0)+(0:1)$);
	\draw (-2, 0) -- ($(-2, 0)+(120:1)$);
	\draw (-2, 0) -- ($(-2, 0)+(240:1)$);
	\draw[fill] (-2, 0) circle  (0.05);
	\draw[ultra thick]  (2, 0) circle (1);
	\draw ($(2, 0)+(45:0.5)$) -- ($(2, 0)+(45:1)$);
	\draw ($(2, 0)+(135:0.5)$) -- ($(2, 0)+(135:1)$);
	\draw ($(2, 0)+(225:0.5)$) -- ($(2, 0)+(225:1)$);
	\draw ($(2, 0)+(315:0.5)$) -- ($(2, 0)+(315:1)$);
	\draw (2, 0) circle (0.5);
	\draw[fill] ($(2, 0)+(315:0.5)$) circle  (0.05);
	\draw[fill] ($(2, 0)+(225:0.5)$) circle  (0.05);
	\draw[fill] ($(2, 0)+(135:0.5)$) circle  (0.05);
	\draw[fill] ($(2, 0)+(45:0.5)$) circle  (0.05);
	\draw[ultra thick] (6,0) circle (1);
	\draw (5, 0) -- (7,0);
	\draw (6,-1) -- (6,1);
	\draw[ultra thick]  (10, 0) circle (1);
	\draw (9, 0) -- ($(10, 0)+(180:0.4)$);
	\draw (10, 0) circle (0.4);
	\draw[fill] (9.6, 0) circle  (0.05);
	
	\end{tikzpicture}
	\caption{Examples of Jacobi diagrams}
	\label{fig:Jacobi_examples}
\end{figure}

\begin{figure}[H]
	\centering
	\begin{tikzpicture}[scale=0.5]
	\coordinate (s) at (0,0);
	\draw[-{Stealth[length=3mm]}, ultra thick] ($(s)+ (45:-2)$) arc(225:315:2);
	\coordinate (t) at (4,0);
	\draw[-{Stealth[length=3mm]}, ultra thick] ($(t)+ (45:-2)$) arc(225:315:2);
	\coordinate (u) at (8,0);
	\draw[-{Stealth[length=3mm]}, ultra thick] ($(u)+ (45:-2)$) arc(225:315:2);
	\coordinate (s1) at ($(s)+(-1, 0)$);
	\coordinate (t1) at ($(t)+(-1, 0)$);
	\coordinate (u1) at ($(u)+(-1, 0)$);
	\coordinate (s2) at ($(s)+(1, 0)$);
	\coordinate (t2) at ($(t)+(1, 0)$);
	\coordinate (u2) at ($(u)+(1, 0)$);
	\draw ($(s)+(0, -2)$) -- ($(s)+(0, -1)$);
	\draw (s1) -- ($(s)+(0, -1)$);
	\draw ($(s)+(0, -1)$) -- (s2);
	\draw[fill] ($(s)+(0, -1)$) circle(0.1);
	
	\draw (t1) -- ($(t)+(70:-2)$);
	\draw ($(t)+(110:-2)$) -- (t2);
	
	\draw (u1) -- ($(u)+(110:-2)$);
	\draw ($(u)+(70:-2)$) -- (u2);
	\node[scale=1.5] at ($0.5*(s) + 0.5*(t)-(0,1)$) {$-$};
	\node[scale=1.5] at ($0.5*(u) + 0.5*(t)-(0, 1)$) {$+$};
	\node[scale=1.5] at ($0.5*(s) - 0.9*(t)-(0,1)$) {$STU = $};
	\end{tikzpicture}
	\label{fig:STU-relation}
\end{figure}
The multiplication is defined in a similar way to the algebra $\hat{\mathcal{A}}_n$. The STU relation imposes additional constraints on an internal graph of Jacobi diagrams:
\begin{figure}[H]
	\vspace{0mm}
	\centering\leavevmode
	\includegraphics[width=10.5cm]{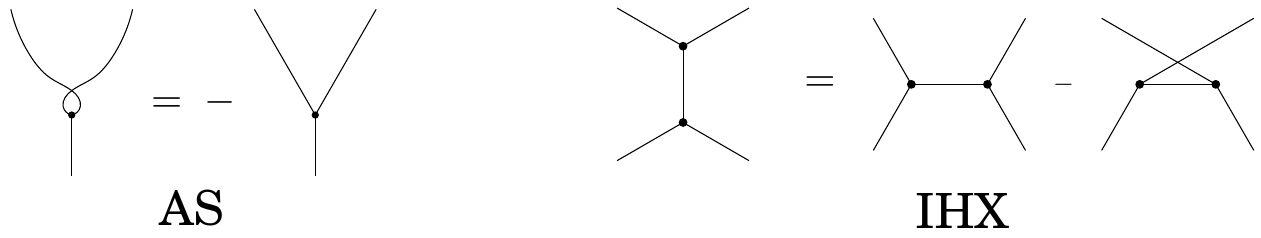}
	\vspace{0mm}
\end{figure}

It is also possible to define a weight system $\varphi^{\mathfrak{g}}_{R}$: with external vertices (i.e. on the distinguished cycle) one associates generators of the Lie algebra $\mathfrak{g}$ in the representation $R$ while with internal vertices one associates generators in the adjoint representation (i.e. structure constants). This construction should be clear from the following example:
\be
\hspace{-35mm}
\varphi^{\mathfrak{g}}_{R} \Big(
\put(12,3){\circle*{24}}
\put(12,3){\color{white}\circle*{21}}
\put(12,3){\circle*{2}}
\put(12,3){\line(0,1){12}}
\put(12,3){\line(2,-1){10.5}}
\put(12,3){\line(-2,-1){10.5}}
\put(25,0){$\Big) = f_{abc} \, \Tr_R  \left( T^aT^bT^c \right).$}
\ee

If we consider the adjoint representation, then the AS and IHX relations act not only on the internal graph, but also on the external (distinguished) cycle. Graphically, we depict such a cycle not with a thick line, symbolizing an arbitrary representation, but with an ordinary thin line, the same as the internal graph.

\bigskip

Since the entire construction contains 4 different algebras, in order not to get confused in their relationships, we have drawn the following diagram. It shows which algebras are responsible for which invariants and how they relate to each other:

\be
\framebox[1\width]{
\begin{tabular}{c}
Chern-Simons\\
knot invariants
\end{tabular}
} \hspace{10mm} 
\begin{tabular}{c}
$\varphi^{\mathfrak{g}}_{R}$\\
$\Longleftarrow$\\
\phantom{a}
\end{tabular}
\hspace{10mm}
\framebox[1\width]{
\begin{tabular}{c}
Kontsevich integral\\
$\sum \limits_{n=0}^{\infty} \hbar^n \sum \limits_{m=1}^{\dim {\mathcal{V}}_n} D_{n,m} \cdot v_{n,m}^{\mathcal{K}}$
\end{tabular}
} \hspace{18mm}  \hspace{7.5mm}
\framebox[1\width]{
\begin{tabular}{c}
Vogel's $\Lambda$ algebra\\
relations $\langle \text{AS, IHX} \rangle$
\end{tabular}
} \hspace{1.5mm}
\nonumber\\
\nearrow \hspace{35mm} \nwarrow \hspace{44mm} \circlearrowleft \hspace{17mm}
\nonumber \\
\hspace{-0mm}
\framebox[0.95\width]{
\begin{tabular}{c}
Knot invariants\\
Reidemeister moves
\end{tabular}
}
 \  \hspace{0.5mm} \Longrightarrow \
\framebox[0.95\width]{
\begin{tabular}{c}
Algebra $\mathcal{V}$ of \\
Vassiliev invariants \\
relations $\langle \text{1T, 4T} \rangle$
\end{tabular}
} \ \Leftarrow \hspace{-0.5mm} \Rightarrow \
\framebox[0.95\width]{
\begin{tabular}{c}
Algebra $\mathcal{A}$ of\\
chord diagrams \\
relations $\langle \text{1T, 4T} \rangle$
\end{tabular}
} \ \Leftarrow \hspace{-0.5mm} \Rightarrow \
\framebox[0.95\width]{
\begin{tabular}{c}
Algebra $\mathcal{C}$ of\\
closed Jacobi diagrams \\
relations $\langle \text{1T, STU} \rangle$
\end{tabular}
}
\nonumber
\ee

\subsubsection{Vogel $\Lambda$-algebra}

$\Lambda$ is a vector space generated by 3-legged diagrams of valence 3 modulo the AS and IHX relations. There is also condition for a diagram in $\Lambda$ to be antisymmetric with respect to permutation of its legs, however, it is unknown whether there are any 3-legged diagrams other than antisymmetric.

There exists a way to multiply any connected trivalent diagram modulo AS+IHX by an element of $\Lambda$. To do so, one should  insert a 3-legged $\Lambda$-diagram into a trivalent vertex of the other diagram. Hence, $\Lambda$ is an algebra with respect to that vertex multiplication and some other spaces of diagrams (like primitive Jacobi diagrams $\mathcal{P}$) acquire the structure of $\Lambda$-module:

\begin{figure}[H]
	\centering
\begin{tikzpicture}[scale=0.7]
\draw (-1.99, 0.32) -- ($(-2, -0.1)+(90:1)$);
\draw (-1.33, -0.8) -- ($(-1.33,-0.8)+(120:1)$);
\draw (-2.17, 0.05) -- ($(-2.17, 0.05)+(240:1)$);
\draw (-2.5, -0.5) -- ($(-2.5, -0.5)+(0:1)$);
\draw (-1.5, -0.5) -- ($(-1.5,-0.5)+(120:1)$);
\draw (-2, 0.35) -- ($(-2, 0.35)+(240:1)$);
\draw[ultra thick]  (0.8, 0) circle (1);
\draw (0.8, 0) -- ($(0.8, 0)+(0:1)$);
\draw (0.8, 0) -- ($(0.8, 0)+(120:1)$);
\draw (0.8, 0) -- ($(0.8, 0)+(240:1)$);
\draw[fill] (0.8, 0) circle  (0.05);
\draw[fill] (-0.8, 0) circle  (0.05);
\put(45,-2){$=$}
\put(-42,-7){$u$}
\put(77,-5){$u$}
\draw[ultra thick]  (4, 0) circle (1);
\draw (4.01, 0.42) -- ($(4, 0)+(90:1)$);
\draw (4.67, -0.7) -- ($(4.67,-0.7)+(120:1)$);
\draw (3.83, 0.15) -- ($(3.83, 0.15)+(240:1)$);
\draw (3.5, -0.4) -- ($(3.5, -0.4)+(0:1)$);
\draw (4.5, -0.4) -- ($(4.5,-0.4)+(120:1)$);
\draw (4, 0.45) -- ($(4, 0.45)+(240:1)$);
\end{tikzpicture}
\end{figure}

$\Lambda$-algebra was extensively studied by P. Vogel \cite{vogel2011algebraic}. Its multiplication is abelian and it was conjectured to be isomorphic to polynomial algebra $R_0 = \mathbb{Q}[t] \oplus \omega \mathbb{Q}[t, \sigma, \omega] $. However, $\Lambda$ turns out to have a divisor of zero: $\hat{t} \cdot \hat{P} = 0$.

\subsubsection{Characters on $\Lambda$}

For $\Lambda$ there is a construction similar to the Lie algebra weight system. For an element of $\Lambda$ structure constant $f_{abc}$ is assigned to each vertex and metric tensor $g_{ab}$ (non-degenerate Killing form) is assigned to each edge:
\be
\Phi_{\mathfrak{g}}: \Lambda \rightarrow L^{\otimes 3}, \
\put(23,3){\line(0,1){15}}
\put(23,3){\line(3,-2){14.5}}
\put(23,3){\line(-3,-2){14.5}}
\put(23,3){\circle*{2}}
\hspace{12mm} \mapsto \ f_{abc}
\nonumber
\ee
Then relation AS means anticommutativity and IHX means Jacobi identity.

After contraction of the duplicate indices we end up with a tensor with 3 free indices. But for simple Lie algebra there is only one invariant tensor of rank 3, namely the structure constant:
\be
\label{char}
\Phi_{\mathfrak{g}}(\hat{x}) = \chi_{\mathfrak{g}}(\hat{x}) \, f_{abc}, \qquad \forall\, \hat{x} \in \Lambda.
\ee
The coefficient of proportionality between the resulting tensor and the structure constant is called a character of the Lie algebra taken on that element of $\Lambda$. The character on $\Lambda$ coming from a Lie algebra $\mathfrak{g}$ defined as above is denoted by $\chi_{\mathfrak{g}}: \Lambda \rightarrow \mathbb{Q}$.

For the elements of $\Lambda$ expressible in the form of a polynomial in $t$, $\sigma$ and $\omega$ (currently, all known elements of $\Lambda$ are of this type) one only needs to know the values that $\chi_{\mathfrak{g}}$ takes on $t$, $\sigma$ and $\omega$. Those are dependent on ${\mathfrak{g}}$ and on the choice of metric in $\mathfrak{g}$. For any renormalization of the metric $t$, $\sigma$ and $\omega$ also rescale as variables of degree 1, 2 and 3 correspondingly. Hence, the values $\chi_{\mathfrak{g}}(\hat{\omega}\hat{\sigma})/\chi_{\mathfrak{g}}(\hat{\omega}\hat{t}^2)$ and $\chi_{\mathfrak{g}}(\omega)/\chi_{\mathfrak{g}}(t^3)$ are invariant and depend only on ${\mathfrak{g}}$.

Instead of variables $t, \sigma, \omega$ there is another rather convenient parameterization via $\alpha, \beta, \gamma$:
\begin{equation}
\label{alphaparamtrization}
	\alpha + \beta +\gamma = t, \quad \alpha \beta + \beta \gamma + \alpha \gamma = \sigma-2t^2, \quad \alpha \beta \gamma = \omega-t\sigma\,.
\end{equation}
The values $\alpha,\ \beta,\ \gamma$ for simple Lie algebras were calculated by Pierre Vogel in \cite{vogel2011algebraic} and are shown in Table \ref{tab:char} in the same normalization as in \cite{mkrtchyan2012casimir}:
\begin{table}[h]
	\centering
	\begin{tabular}{|c|c|c|c|}
		\hline
		Lie algebra $\mathfrak{g}$& $\alpha$ & $\beta$ & $\gamma$ \\
		\hline
		$\mathfrak{sl}_{N}$ & $-2$ & $2$ &$N$ \\
		\hline
		$\mathfrak{so}_{N}$ & $-2$ & $4$ & $N-4$\\
		\hline
		$\mathfrak{sp}_{2N}$ & $-2$ & $1$ &$N+2$ \\
		\hline
		$G_2$ & $-2$ & $10/3$ & $8/3$ \\
		\hline
		$F_4$ & $-2$ & $5$ &$6$ \\
		\hline
		$E_6$ & $-2$ & $6$ & $8$ \\
		\hline
		$E_7$ & $-2$ & $8$ & $12$\\
		\hline
		$E_8$ & $-2$ & $12$ & $20$ \\
		\hline
	\end{tabular}
	\caption{Vogel parameters.}	
	\label{tab:char}
\end{table}

Since variables $t, \sigma, \omega$ are symmetric in $\alpha, \beta, \gamma$, what obviously follows from \eqref{alphaparamtrization}, then characters $\chi_{\mathfrak{g}}(\hat{x})$ are also symmetric polynomials in $\alpha, \beta, \gamma$.

\bigskip

Now let us ask the following question. Does there exist such element of $\Lambda$-algebra whose character vanishes for the $\mathfrak{g}$ algebra?
\be
\chi_{\mathfrak{g}}( \hat{x}_{\mathfrak{g}} ) &\sim& P_{\mathfrak{g}}, \qquad \text{where} \\
P_{\mathfrak{sl}} &=& (\alpha+\beta)\,(\beta+\gamma)\,(\alpha+\gamma), \hspace{0mm} \nonumber \\
P_{\mathfrak{osp}} &=& (\alpha+2\beta)(2\alpha+\beta)\,(\beta+2\gamma)(2\beta+\gamma)\,(\alpha+2\gamma)(2\alpha+\gamma),  \nonumber \\
P_{\mathfrak{exc}} &=& (\alpha-2\beta-2\gamma)\,(\beta-2\alpha-2\gamma)\,(\gamma-2\alpha-2\beta). \nonumber 
\ee

To answer this question one should rewrite these polynomials in terms of $t, \sigma, \omega$ and use the unique graded homomorphism
$\psi : \mathbb{Q}[t] \oplus \omega\,\mathbb{Q}[t,\sigma,\omega]\rightarrow\Lambda$:
\be
\psi\left( t \right) = \hat{t}, \quad
\psi\left( \omega \right) = \hat{\omega}, \quad
\psi\left( \sigma^p\omega \right) = \hat{\omega}_p,
\ee
where the diagram $\hat{\omega}_p, \ p\geq0$ can be expressed in terms of diagrams $\hat{x}_n$:
\begin{figure}[!h]
	\centering
	\begin{tikzpicture}		
		\node at (3.7, 0.2) {$\hat{x}_n=$};
		\draw (4., -0.5) -- (5.4, -0.5);
		\draw (4.7, 0.6) -- +(0, 0.4);
		\draw (4.7, 0.3) circle (0.3);
		\draw ($(4.7, 0.3)+(50:-0.3)$) -- (4.2, -0.5);
		\draw ($(4.7, 0.3)+(60:-0.3)$) -- (4.3, -0.5);
		\draw ($(4.7, 0.3)+(70:-0.3)$) -- (4.4, -0.5);
		\draw ($(4.7, 0.3)+(80:-0.3)$) -- (4.5, -0.5);
		\node at (4.7, -0.3) {$\cdot$};
		\node at (4.8, -0.3) {$\cdot$};
		\node at (4.9, -0.3) {$\cdot$};
		\draw ($(4.7, 0.3)+(120:-0.3)$) -- (5.1, -0.5);
		\draw ($(4.7, 0.3)+(130:-0.3)$) -- (5.2, -0.5);
	\end{tikzpicture} 
	\label{fig:some-of-Lambda}
\end{figure}

For example, $
\psi\left( \omega P_{sl} \right) = \widehat{\omega P_{sl}} \equiv \frac{4}{3}\hat{t}\hat{x_5} + \frac{16}{9}\hat{t}^6 - \frac{8}{3}\hat{t}^3\hat{x_3} - \frac{4}{9}\hat{x}_3^2.
$




\subsubsection{Zero divisor}
\label{zero_divisor}
We have already mentioned that $\Lambda$-algebra has a divisor of zero $\hat{t} \cdot \hat{P} = 0$ and now we describe it. A more detailed description with all the proofs can be found in \cite[Theorem 8.4]{vogel2011algebraic}. Let $\hat{U}$ be the following 6-legged diagram:

\begin{picture}(300,120)(-150,-5)
\put(0,50){$\hat{\text{U}} \ \, = $}

\put(50,53){\line(0,1){17}}
\put(80,53){\line(0,1){17}}
\put(50,70){\line(5,3){15}}
\put(65,79){\line(5,-3){15}}
\put(50,53){\line(5,-3){15}}

\put(110,53){\line(0,1){17}}
\put(80,70){\line(5,3){15}}
\put(95,79){\line(5,-3){15}}
\put(95,44){\line(5,3){15}}

\put(65,27){\line(0,1){17}}
\put(95,27){\line(0,1){17}}
\put(65,44){\line(5,3){15}}
\put(80,53){\line(5,-3){15}}
\put(80,18){\line(5,3){15}}
\put(65,27){\line(5,-3){15}}

\put(50,36){\line(0,1){17}}
\put(50,36){\line(5,-3){15}}

\put(110,36){\line(0,1){17}}
\put(95,27){\line(5,3){15}}

\put(80,88){\line(5,-3){15}}
\put(65,79){\line(5,3){15}}

\put(80,88){\line(0,1){12}}
\put(80,6){\line(0,1){12}}
\put(50,70){\line(-2,1){15}}
\put(50,36){\line(-2,-1){15}}
\put(110,36){\line(2,-1){15}}
\put(110,70){\line(2,1){15}}

\end{picture}

Then we  \text{define}  $\hat{P} \neq 0 \in \Lambda$ by  removing a neighborhood of a trivalent  vertex from $V$:

\begin{picture}(850,30)(-375,-25)
\put(-314,-22){V= {\Large \text{$\frac1{6!}\sum \limits_{\sigma \in S_6}$}} $ \text{sign}(\sigma) \ \cdot$}
\put(-199,-20){\circle{30}}
\put(-203,-22){$\hat{\text{U}}$}

\put(-197,-35){\line(1,0){32}}
\put(-187,-29){\line(1,0){22}}
\put(-184.5,-23){\line(1,0){19.5}}
\put(-184.5,-17){\line(1,0){19.5}}
\put(-187,-11){\line(1,0){22}}
\put(-197,-5){\line(1,0){32}}

\put(-165,-40){\line(0,1){40}}
\put(-165,-40){\line(1,0){20}}
\put(-165,-0){\line(1,0){20}}
\put(-145,-40){\line(0,1){40}}
\put(-157,-22){$\sigma$}

\put(-145,-35){\line(1,0){32}}
\put(-145,-29){\line(1,0){22}}
\put(-145,-23){\line(1,0){19.5}}
\put(-145,-17){\line(1,0){19.5}}
\put(-145,-11){\line(1,0){22}}
\put(-145,-5){\line(1,0){32}}

\put(-111,-20){\circle{30}}
\put(-114,-22){$\hat{\text{U}}$}

\put(-90,-22){$=$}
\put(-60.7,-20){\circle{25}}
\put(-65,-22){$\hat{P}$}
\put(-48,-20){\line(1,0){14.5}}
\put(-53.3,-10){\line(2,-1){20}}
\put(-53.3,-30){\line(2,1){20}}
\put(-34,-20){\circle*{2}}
\qbezier[8](-34,-15)(-29,-15)(-29,-20)
\qbezier[8](-29,-20)(-29,-25)(-34,-25)
\qbezier[8](-34,-25)(-39,-25)(-39,-20)
\qbezier[8](-39,-20)(-39,-15)(-34,-15)
\put(-20,-15){$\footnotesize \text{remove}$}
\put(-21,-15){\vector(-2,-1){10}}
\end{picture}

\vspace{0mm}
\be
\psi(\omega P_{sl}P_{osp}P_{exc}) = - 2^{10} \hat{P} \nonumber \\
\hat{t} \cdot \hat{P} = 0 \nonumber, \quad \hat{t} = 
\put(10,0){\line(1,0){16}}
\put(10,0){\line(1,2){8}}
\put(18,16){\line(1,-2){8}}
\put(18,16){\line(0,1){8}}
\put(10,0){\line(-2,-1){8}}
\put(26,0){\line(2,-1){8}}
\ee

\subsubsection{Peculiar $D_{17}$ diagram} 


Now we are ready to answer the question \eqref{question}.  Indeed, $\hat{P} = \psi \left( \omega \, P_{sl} \, P_{osp} \, P_{exc} \right)$ is an element of degree $15$ in $\Lambda$ that is killed by characters of either $\mathfrak{sl}_n$, $\mathfrak{so}_n$, $\mathfrak{sp}_{2n}$ or exceptional Lie algebra series. It can be multiplied on a "Mercedez-Benz" Jacobi diagram of degree 2.  Finally, we end  up with the $D_{17}$ Jacobi diagram that is undetected by all weight systems coming from simple Lie (super)algebras:
\be
\label{weight17}
\hspace{-35mm} D_{17} := \hat{P} \cdot 
\put(12,3){\circle*{24}}
\put(12,3){\color{white}\circle*{21}}
\put(12,3){\circle*{2}}
\put(12,3){\line(0,1){12}}
\put(12,3){\line(2,-1){10.5}}
\put(12,3){\line(-2,-1){10.5}} \qquad \quad \  \nonumber \\ \\
\Phi_{\mathfrak{g}}\left(D_{17}\right) = \Phi_{\mathfrak{g}}\left( \hat{P} \cdot\hat{t}  \right) \cdot \Phi_{\mathfrak{g}}\Big( 
\put(12,3){\circle*{24}} 
\put(12,3){\color{white}\circle*{21}}
\put(0,3){\line(1,0){24}}
\put(24,0){$\Big)$} 
\put(32,0){$=0$}
\nonumber
\ee

Let us stress that $D_{17} \neq 0$. It follows from two statements. The first one is that $\hat{P} \neq 0$, because $\chi_{D(2,1;\lambda)}(\hat{P}) \neq  0$. The second one is Corollary 4.7 from Vogel's paper \cite{vogel2011algebraic}, which states 
$\put(0,0){$\hat{x} \,\cdot $} 
\put(20,3){\circle*{16}}
\put(20,3){\color{white}\circle*{14}}
\put(20,3){\circle*{2}}
\put(20,3){\line(0,1){8}}
\put(20,3){\line(2,-1){6.5}}
\put(20,3){\line(-2,-1){6.5}}
\put(33,0){$\neq 0$}$ \qquad \qquad \quad for any nonzero $\hat{x} \in \Lambda$.

\subsubsection{Adjoint representation} Another interesting property of $D_{17}$ is that it vanishes by restricting to the adjoint representation, what is a consequence of $\hat{t} \cdot \hat{P} =0$ identity in $\Lambda$. It follows from the fact that for the adjoint representation the distinguished cycle of the Jacobi diagram becomes ordinary edges and one can factor out $\hat{t}$ from the "Mercedez-Benz" diagram.

\medskip

\hspace{-5mm}Furthermore, consider diagram $V$ (of 16 degree) and present it in the following form
\be
\put(-25,0){$V = \hat{P} \cdot$}
\put(20,3){\circle{20}}
\put(10,3){\line(1,0){20}}
\put(10,3){\circle*{2}}
\put(30,3){\circle*{2} ,}
\ee
where the circle trivalent diagram in the right hand side has no distinguished cycle. Such diagram is suitable for the adjoint representation as we discussed in Section \ref{Jacobialg}. Therefore, applying the weight system we get 
\be
\Phi_{\mathfrak{g}} \Big(
\put(10,3){\circle{20}}
\put(0,3){\line(1,0){20}}
\put(0,3){\circle*{2}}
\put(20,3){\circle*{2}}
\put(21,0){$\Big)$} \hspace{10mm} = 2t \cdot \Phi_{\mathfrak{g}} \Big( 
\put(10,3){\circle{20}}
\put(21,0){$\Big)$} \hspace{10mm} = 2\,t\,\dim_{adj} \ \Rightarrow \nonumber \\
\Rightarrow \ \Phi_{\mathfrak{g}}(V) = 2\, \Phi_{\mathfrak{g}}\left(\hat{P} \cdot \hat{t}\right) \cdot \dim_{adj} = 0. \hspace{12mm}
\ee

Thus, the diagram $V$  is undetected by the adjoint representation. It means that the contribution of this diagram vanishes into the universal knot polynomials whereas in the Kontsevich integral it is present.

\section{Implications from Vogel's results}
\label{sec:3}
In this section we discuss several implications of Vogel's results. We start with what questions and problems arise in Chern-Simons theory and how they can be solved. This is the subject of the first two subsections. In the third subsection we discuss that inverting Vogel's logic one can start from knot invariants and arrive at a definition of Lie algebras and their classification in a fairly natural way.

\subsection{Chern-Simons theory}
\label{1imp}
In Section \ref{CSKI} we discussed that each group factor $G_{n,m}^{\mathfrak{g},R}$ that occurs in the perturbative decomposition \eqref{wilsonexpansion} of the Wilson loop in Chern-Simons theory is a value of a weight system of the corresponding Lie algebra on a suitable chord diagram or on a linear combination of chord diagrams: 
\be
G_{n,m}^{\mathfrak{g},R} = \varphi^{\mathfrak{g}}_R \left( D_{n,m} \right).
\ee
Vogel's theorem states that there exist primitive (with respect to the comultiplication in the algebra of chord diagrams) chord diagrams that vanish under the action of the weight system for any semisimple (super) Lie algebra $\mathfrak{g}$ and its finite-dimensional representation $R$, for example \eqref{weight17}: 
\be
\varphi^{\mathfrak{g}}_R \left( D_{17} \right) = 0. 
\ee
Therefore, in Chern-Simons theory with any finite-dimensional gauge Lie algebra the corresponding group factors vanish. \textbf{It means that the corresponding Vassiliev invariants cannot be extracted from Wilson loops.}

We postpone the discussion of infinite-dimensional Lie algebras and/or infinite-dimensional representations until the next subsection \eqref{2imp}, and now we discuss only such consequences, which correspond to finite-dimensional case.

The dramatic nature of the conclusions drawn from this result depends on \textit{two forks}. \textit{The first one} is whether Vassiliev invariants are a complete knot invariant or not. \textit{The second one} is whether the Vassiliev invariants are an overdetermined system or not, i.e. whether there are relations between the primitive Vassiliev invariants or not. 

$\bullet$ If the Vassiliev invariants form a complete knot invariant and there are no hidden relations between them, then there exist knots that Wilson loops are unable to distinguish. This begs the question whether there are other gauge invariants in Chern-Simons theory that do not reduce to Wilson loops.

$\bullet$ If Vassiliev's invariants do not form a complete knot invariant or if there are hidden relations between them, then no dramatic conclusions can be made for now. In this case, further research in this direction is required.

\subsection{Infinite-dimensional case}
\label{2imp}

$\bullet$ A promising direction for further research is the case of infinite-dimensional Lie algebras. However, an algebra is needed in which there will be a non-degenerate metric, so that it will be possible to contract indices and anti-symmetrize the structure constants $f^{ab}_c$ over all three indices $a,b,c$. In this case, such an algebra will still give rise to the weight system of the Vogel algebra, but some arguments from the proof of the theorem may no longer work, for example, the first equality in formula \eqref{weight17}. In particular, one may seek for a such Lie algebra $\mathfrak{g}$ that the following hold:
\be
\Phi_{\mathfrak{g}} \Big( 
\put(12,3){\circle*{24}}
\put(12,3){\color{white}\circle*{21}}
\put(12,3){\circle*{2}}
\put(12,3){\line(0,1){12}}
\put(12,3){\line(2,-1){10.5}}
\put(12,3){\line(-2,-1){10.5}} 
\put(25,0){$\Big)$} 
\qquad \quad \neq \Phi_{\mathfrak{g}}\left( \hat{t}  \right) \cdot \Phi_{\mathfrak{g}}\Big( 
\put(12,3){\circle*{24}} 
\put(12,3){\color{white}\circle*{21}}
\put(0,3){\line(1,0){24}}
\put(24,0){$\Big)$} 
\ee
This apparently requires Lie algebra with two non-degenerate invariant bilinear forms, one of which is the Killing form.

Another possibility is the presence of two or more invariant 3-tensors. In this case, it will no longer be possible to determine the character of \eqref{char} and the contribution from the $\Lambda$-algebra in the diagram $D_{17}$ will simply no longer be factorable.

\bigskip

\hspace{-5mm}$\bullet$ The direction associated with infinite-dimensional representations seems much less promising. However, strictly speaking, it also needs to be carefully analyzed. The idea is that the trace of the infinite-dimensional representation gives an infinitely large contribution, which compensates for the zero from the zero divisor in formula (\ref{weight17}):
\be
\Phi_{\mathfrak{g}}\left(D_{17}\right) = \Phi_{\mathfrak{g}}\left( \hat{P} \cdot\hat{t}  \right) \cdot \Phi_{\mathfrak{g}}\Big( 
\put(12,3){\circle*{24}} 
\put(12,3){\color{white}\circle*{21}}
\put(0,3){\line(1,0){24}}
\put(24,0){$\Big)$} 
\put(32,0){$=0\cdot \infty = \text{finite}$}
\nonumber
\ee
However, in this case this diagram could still be suppressed compared to the contributions from other diagrams that do not have a zero divisor.


\subsection{Alternative axiomatization of Lie algebras}
\label{4imp}

\subsubsection{Reidemeister implies 4T-relation}
Consider the expression in Figure \ref{fig:4T-knot}, where $v$ is an arbitrary Vassiliev invariant and singular knots where it is evaluated coincide outside of the depicted zone.

\begin{figure}[H]
	\centering
		\resizebox*{.75\linewidth}{1.5cm}
	{\begin{tikzpicture}[scale=0.8, use Hobby shortcut]
		\coordinate (a) at (-2, 0);
		\coordinate (b) at (2, 0);
		\coordinate (c) at (6, 0);
		\coordinate (d) at (10, 0);
		\node[scale=2.] at ($(a)-(1.4, 0)$) {$v($};
		\node[scale=2.] at ($(a)+(1.12, 0)$) {$)$};
		
		\node[scale=2.] at ($(b)-(1.74, 0)$) {$-v($};
		\node[scale=2.] at ($(b)+(1.12, 0)$) {$)$};
		
		\node[scale=2.] at ($(c)-(1.74, 0)$) {$-v($};
		\node[scale=2.] at ($(c)+(1.12, 0)$) {$)$};
		
		\node[scale=2.] at ($(d)-(1.74, 0)$) {$+v($};
		\node[scale=2.] at ($(d)+(2, 0)$) {$)=0$};
		\draw[dashed] (b) circle (1);
		
		\draw[-{Stealth[length=2.5mm]}, thick] ($(b)-(1, 0)$)--($(b)+(1,0)$);
		\draw[-{Stealth[length=2.5mm]}, thick] ($(b)+(45:0.5)$)--($(b)+(45:1)$);
		\draw[thick] ($(b)+(45:-1)$)--($(b)+(45:0.3)$);
		\draw [fill] ($(b)+(0:0.5)$) circle (0.075);
		\draw [fill] ($(b)$) circle (0.075);
		\draw[-{Stealth[length=2.5mm]}, thick] ($(b)+(90:-1)$)..($(b)+(135:-0.5)$)..($(b)+(0:0.5)$) ..($(b)+(45:0.4)$)..($(b)+(90:1)$);
		
		\draw[dashed] (a) circle (1);
		
		\draw[-{Stealth[length=2.5mm]}, thick] ($(a)-(1, 0)$)--($(a)+(1,0)$);
		\draw[-{Stealth[length=2.5mm]}, thick] ($(a)+(45:-1)$)--($(a)+(45:1)$);
		\draw [fill] ($(a)+(0:-0.5)$) circle (0.075);
		\draw [fill] ($(a)$) circle (0.075);
		\draw[thick] ($(a)+(90:-1)$)..($(a)+(85:-0.8)$)..($(a)+(55:-0.5)$);
		\draw[-{Stealth[length=2.5mm]}, thick] ($(a)+(35:-0.5)$)..($(a)+(0:-0.5)$)..($(a)+(135:0.5)$) ..($(a)+(90:1)$);

		\draw[dashed] (c) circle (1);
		
		\draw[-{Stealth[length=2.5mm]}, thick] ($(c)-(1, 0)$)--($(c)+(1,0)$);
		\draw[-{Stealth[length=2.5mm]}, thick] ($(c)+(45:-1)$)--($(c)+(45:1)$);
		\draw [fill] ($(c)+(45:0.4)$) circle (0.075);
		\draw [fill] ($(c)$) circle (0.075);
		\draw[thick] ($(c)+(90:-1)$)..($(c)+(135:-0.4)$)..($(c)+(170:-0.4)$);
		\draw[-{Stealth[length=2.5mm]}, thick] ($(c)+(10:0.4)$)..($(c)+(45:0.4)$)..($(c)+(85:0.75)$)..($(c)+(90:1)$);

		\draw[dashed] (d) circle (1);
		
		\draw[-{Stealth[length=2.5mm]}, thick] ($(d)-(0.3, 0)$)--($(d)+(1,0)$);
		\draw[ thick] ($(d)-(1, 0)$)--($(d)+(-0.5,0)$);
		\draw[-{Stealth[length=2.5mm]}, thick] ($(d)+(45:-1)$)--($(d)+(45:1)$);
		\draw [fill] ($(d)+(45:-0.4)$) circle (0.075);
		\draw [fill] (d) circle (0.075);
		\draw[-{Stealth[length=2.5mm]}, thick] ($(d)+(90:-1)$)..($(d)+(45:-0.4)$)..($(d)+(0:-0.4)$)..($(d)+(100:0.8)$)..($(d)+(90:1)$);
		
	\end{tikzpicture}}
	\caption{Four terms knot relation}
	\label{fig:4T-knot}
\end{figure}

If we resolve double points according to Vassiliev skein relation (see Section \ref{CSKI}), then all terms are reduced due to the third Reidemeister move, which becomes Yang-Baxter equation in algebraic terms. This results in the four-term relation in Figure \ref{fig:4T-knot}.

From this relation on Vassiliev invariants there follows a similar four-term relation on the functions  originating from Vassiliev invariants (Fig. \ref{fig:4T-fun}):

\begin{figure}[!h]
	\centering
			\resizebox*{.75\linewidth}{1.5cm}
	{
		\begin{tikzpicture}[scale=0.8]
			\coordinate (a) at (-2, 0);
			\coordinate (b) at (2, 0);
			\coordinate (c) at (6, 0);
			\coordinate (d) at (10, 0);
			\node[scale=2] at ($(a)-(1.4, 0)$) {$f($};
			\node[scale=2] at ($(a)+(1.12, 0)$) {$)$};
			
			\node[scale=2] at ($(b)-(1.8, 0)$) {$-f($};
			\node[scale=2] at ($(b)+(1.12, 0)$) {$)$};
			
			\node[scale=2] at ($(c)-(1.8, 0)$) {$-f($};
			\node[scale=2] at ($(c)+(1.12, 0)$) {$)$};
			
			\node[scale=2] at ($(d)-(1.8, 0)$) {$+f($};
			\node[scale=2] at ($(d)+(2.07, 0)$) {$)=0$};
			\draw[dashed] (a) circle (1);
			\draw[thick] ($(a)+(240:1)$) arc(240:300:1);
			\draw[thick] ($(a)+(0:1)$) arc(0:60:1);
			\draw[thick] ($(a)+(140:1)$) arc(140:180:1);
			\draw ($(a)+(255:1)$)--($(a)+(160:1)$);
			\draw ($(a)+(280:1)$)--($(a)+(20:1)$);
			\draw[dashed] (b) circle (1);
			\draw[thick] ($(b)+(240:1)$) arc(240:300:1);
			\draw[thick] ($(b)+(0:1)$) arc(0:60:1);
			\draw[thick] ($(b)+(140:1)$) arc(140:180:1);
			\draw ($(b)+(285:1)$)--($(b)+(160:1)$);
			\draw ($(b)+(255:1)$)--($(b)+(20:1)$);
			\draw[dashed] (c) circle (1);
			\draw[thick] ($(c)+(240:1)$) arc(240:300:1);
			\draw[thick] ($(c)+(0:1)$) arc(0:60:1);
			\draw[thick] ($(c)+(140:1)$) arc(140:180:1);
			\draw ($(c)+(45:1)$)--($(c)+(160:1)$);
			\draw ($(c)+(270:1)$)--($(c)+(20:1)$);
			\draw[dashed] (d) circle (1);
			\draw[thick] ($(d)+(240:1)$) arc(240:300:1);
			\draw[thick] ($(d)+(0:1)$) arc(0:60:1);
			\draw[thick] ($(d)+(140:1)$) arc(140:180:1);
			\draw ($(d)+(20:1)$)--($(d)+(160:1)$);
			\draw ($(d)+(270:1)$)--($(d)+(45:1)$);
	\end{tikzpicture}}
	\caption{4-term relation on functions coming from Vassiliev invariants}
	\label{fig:4T-fun}
\end{figure}

\subsubsection{4T = STU $\rightarrow$ AS + IHX = Lie algebras}

It is quite easy to prove that the space of chord diagrams modulo the 4T relation is isomorphic to the space of Jacobi diagrams modulo the STU relation \cite{chmutov2012introduction}.

STU relation itself implies AS and IHX relation on the internal graphs of Jacobi diagrams. These relations are exactly the antisymmetry of Lie bracket and the Jacobi identity. Although these diagrams correspond to a Lie algebra with a non-degenerate invariant metric.

Hence, \textbf{the concept of Lie algebra may be derived from topological knots}.

\subsubsection{Towards the classification of (quasi)simple Lie algebras}
To classify Lie algebras in this approach (using Vogel's $\Lambda$) additional idea of quasisimplicity is needed. Essentially it means that there should be a single invariant antisymmetric rank 3 tensor (the structure constant). This requirement is essential for the concept of Lie algebra character on $\Lambda$ to make sense.

In $\Lambda$ there is a following identity (it is already described in Section \ref{zero_divisor}):
\begin{equation}
	\hat{t} \cdot\hat{P} = 0
\end{equation}

Once $\hat{P}$ is represented as polynomial in $(t, \sigma, \omega)$, it can be factorized into a product of the form:

\begin{equation}
	t \omega (2 t \sigma - \omega - 2 t^3)(27 \omega^2-72 t \sigma \omega +40 t^3 \omega+4 \sigma^3+29t^2 \sigma^2 - 24t^4 \sigma) (27 \omega-45 t \sigma +40 t^3)  =0
\end{equation}

Applying $\chi_L$, which is a ring homomorphism from $\Lambda$ to $\mathbb{Q}$, yields $\chi_L(\hat{t}) \chi_L(\hat{P}) =0$. Hence, for any $L$ either $\chi_L(\hat{t})=0$ or $\chi_L(\hat{P})=0$. In case of the former $L$ is the superalgebra $D_{2, 1, \lambda}$. In case of the latter we may proceed as follows.

Any (quasi)simple Lie algebra that is not $D_{2, 1, \lambda}$ vanishes on $\omega$ or on some of the polynomials $2 t \sigma - \omega - 2 t^3$, $27 \omega^2-72 t \sigma \omega +40 t^3 \omega+4 \sigma^3+29t^2 \sigma^2 - 24t^4 \sigma$ or $27 \omega-45 t \sigma +40 t^3$. If $\chi_L(\omega)=0 $ then $L$ is $\mathfrak{sl}_2$. In case of $\chi_L(2 t \sigma - \omega - 2 t^3)=0$ $L$ is of $\mathfrak{sl}$ type. In case of $\chi_L(27 \omega^2-72 t \sigma \omega +40 t^3 \omega+4 \sigma^3+29t^2 \sigma^2 - 24t^4 \sigma)=0$ $L$ is of $\mathfrak{osp}$ type. The case $\chi_L(27 \omega-45 t \sigma +40 t^3)=0$ corresponds to the exceptional Lie algebras ($\mathfrak{g}_2$/$\mathfrak{f}_4$/$\mathfrak{e}_6$/$\mathfrak{e}_7$/$\mathfrak{e}_8$).

This can be reformulated as follows. Presence of zero divisors in $\Lambda$ puts constraint on the possible values of Vogel parameters, leaving out only a few continuous one-parameter families of Vogel parameters that correspond to some Lie algebras. 

Another way to classify Lie algebras using Vogel theory was suggested by R. Mkrtchyan \cite{mkrtchyan2016road}. He analyzed the universal formula for quantum dimension (\ref{eq:qdim}). This formula corresponds to the Kontsevich integral of the unknot taken in the adjoint representation and expressed in terms of Vogel parameters \cite{bar2000wheels}:

\begin{equation}\label{eq:qdim}
	\text{qdim}=\frac{\sinh((\alpha-2t)x/4) \sinh((\beta-2t)x/4) \sinh((\gamma-2t)x/4)}{\sinh(\alpha x /4) \sinh(\beta x/4) \sinh(\gamma x /4)}
\end{equation}

R. Mkrtchyan required this formula to have no poles, i.e. zeroes of the denominator have to be cancelled by zeroes of the numerator.  The requirement is valid for finite dimensional Lie algebras, but it may leave out infinite dimensional case. From this condition he found that possible Vogel parameters organize into several series, which is also sort of an alternative classification of Lie algebras. What was left out of considerations by R.Mkrtchyan is the presence of zero divisors in $\Lambda$.

One can combine the results of Mkrtchyan with the aforementioned zero divisors consequence. These are the points in Vogel's plane that satisfy both $t P =0$ and make the formula for the quantum dimension  (\ref{eq:qdim}) to be a regular function of $x$. It leaves out only a few possible values for the Vogel parameters, which are listed in Table \ref{tab:Vog-classification} in Mkrtchyan's notation:

\begin{table}[h]
	
	\centering
	\begin{tabular}{|c|c|c|c|}
		
		\hline
		$\alpha$ & $\beta$ & $\gamma$ & $\mathfrak{g}$ \\
		\hline
		$-\lambda-1$& $1$ & $\lambda$ & $D_{2,1,\lambda}$ \\
		\hline
		$2$ &$-2$ &$n$& $\mathfrak{su}_n$ \\
		\hline
		$4$&$-2$&$n-4$& $\mathfrak{so}_n$ \\
		\hline
		\multicolumn{3}{|c|}{$\alpha + \beta + 2\gamma =0$} & $\mathfrak{sl}_2$ \\
		\hline
		$3$ &$-5$&$-4$& $\mathfrak{g}_2$ \\
		\hline
		$2$ & $-5 $&$-6$& $\mathfrak{f}_4$ \\
		\hline
		$-3$ & $-4$ & $1$ & $\mathfrak{e}_6$ \\
		\hline
		$-6$ & $-4$ & $1$ & $\mathfrak{e}_7$ \\
		\hline
		$-8$ & $-5$ & $1$ & $\mathfrak{e}_{7\text\textonehalf}$  \\
		\hline
		$-6$ & $-10$ & $1$ & $\mathfrak{e}_8$ \\
		\hline

	\end{tabular}
	\qquad
	\begin{tabular}{|c|c|c|c|}
		\hline
		$\alpha$ & $\beta$ & $\gamma$ & $\mathfrak{g}$ \\
		\hline
		$4$ & $1$ & $1$ & $Y_{31}$ \\
		\hline
		$6$ & $5$ & $22$ & $Y_{32}$ \\
		\hline
		$18$ & $4$ & $5$ & $Y_{33}$ \\
		\hline
		$14$ & $4$ & $3$ & $Y_{34}$ \\
		\hline
		$2$ & $3$ & $10$ & $Y_{35}$ \\
		\hline
		$3$ & $5$ & $16$ & $Y_{37}$ \\
		\hline
		$1$ & $2$ & $6$ & $Y_{38}$ \\
		\hline
		$2$ & $5$ & $14$ & $Y_{41}$ \\
		\hline
		$1$ & $3$ & $8$ & $Y_{43}$ \\
		\hline
		$10$ & $4$ & $1$ & $Y_{45}$ \\
		\hline
		$12$ & $1$ & $5$ & $Y_{46}$ \\
		\hline
		$1$ & $6$ & $14$ & $Y_{47}$ \\
		\hline
	\end{tabular}
	\caption{Vogel parameters for algebra $\mathfrak{g}$}
	\label{tab:Vog-classification}
\end{table}

We end up with all of the algebras in the standard Dynkin classification, and several additional parameters in the exceptional line. It is unknown whether these values of parameters correspond to any Lie (super)algebra or not, apart from the $\mathfrak{e}_{7\text\textonehalf}$ case which was first found by \cite{mathur1988classification, deligne1996serie, cohen1996computational} and is relatively well studied \cite{landsberg2006sextonions}. The parameters $Y_i$ in the right side of the table yield negative integer numbers when plugged into the dimension formula \eqref{eq:qdim}. This is why Mkrtchyan called them unphysical. Although, the universal dimension formula is in fact the superdimension formula, hence these $Y_i$ may correspond to some Lie superalgebras.

\section{Summary}
\label{sec:4}
Vogel's result has rekindled interest in the perturbative sector of Chern-Simons theory. Usually in QFT non-perturbative quantities (like instantons) carry (much) more information about the original physical system than perturbative ones. For the Chern-Simons theory Vogel's result can lead (if a number of other conditions are met) to the exact opposite case: \textbf{Perturbative} Vassiliev invariants may be richer than \textbf{non-perturbative} colored HOMFLY-PT, Kauffman etc. polynomials. This puzzle is waiting to be solved. The obvious possibilities are:
\begin{itemize}
\item{} There can be Lie algebras with degenerate and/or non-unique metrics -- then 3-valent and Jacobi diagrams are ill-defined and require additional specification.

\item{} Lie algebras can be nice, but infinite dimensional. The most obvious examples are affine Kac-Moody $\hat G$ or continuation of $E$-series to $E_9, E_{10}, \ldots$ Naively 3-valent and Jacobi diagrams now involve infinite sums, still there can be ways to define them.
\end{itemize}

However, it is important to emphasize that today it cannot be claimed that we can distinguish more knots using Vassiliev invariants than using colored HOMFLY-PT, Kauffman, etc. polynomials. This is explained by the fact that all this story takes into account only polynomial relations between Vassiliev invariants, whereas the relations between them can be much more complicated. So it is possible that the state space determined by Wilson loops is exactly the same as that of Vassiliev invariants.

Another very important result is a universal description of the adjoint representation of Lie algebras. This description is given using a diagrammatic technique that greatly simplifies the calculations. It is applicable, at a minimum, to any pure Yang-Mills theory. In algebraic language, the diagrammatic technique is described by an Vogel's $\Lambda$-algebra. This is a commutative algebra, similar to the algebra of polynomials in three variables, but in which there is zero divisor. In Vogel's theory, classical Lie algebras are not distinguished \textit{a priori} in any way, but their locus miraculously turns out to be exactly this zero divisor. This opens a possibility to treat the existence of classical Lie algebras as a consequence of knot invariance, i.e. one can provide an implication
 
 \be
 \boxed{
 	{\rm Reidemeister\ invariance} \ \stackrel{?}{\Longrightarrow} \  {\rm Cartan-Dynkin \ classification}
 }
 \label{implic}
 \ee

\section*{Acknowledgements}

This work was funded by the RSF grant No.24-12-00178.

\printbibliography

\end{document}